\documentclass[aps,prl,reprint,superscriptaddress]{revtex4-2}
\usepackage{amsmath,amssymb}
\usepackage{mathastext}
\usepackage{graphicx}
\usepackage{dcolumn}
\usepackage{bm}
\usepackage{anysize}
\usepackage{epstopdf}
\usepackage{epsfig}
\usepackage[outercaption]{sidecap}
\usepackage[pass,letterpaper]{geometry}
\usepackage{color,colortbl} 
\usepackage{boxhandler}
\usepackage{booktabs}
\usepackage[table,x11names,dvipsnames,table]{xcolor}

\begin{document}

\title{Carrier screening controls transport in conjugated polymers at high doping concentrations}



\author{Muhamed Duhand\v{z}i\'{c}}
\affiliation{Materials Science and Engineering, University of Utah, Salt Lake City, 84112 UT, USA}

\author{Michael Lu-D\`iaz}
\author{Subhayan Samanta}
\author{Dhandapani Venkataraman}
\affiliation{Department of Chemistry, University of Massachusetts Amherst, Amherst, 01003 MA, USA}

\author{Zlatan Ak\v{s}amija}
\email[]{zlatan.aksamija@utah.edu}
\homepage[]{https://nanoenergy.mse.utah.edu/  \\}
\affiliation{Materials Science and Engineering, University of Utah, Salt Lake City, 84112 UT, USA}

\date{\today}
\begin{abstract}
Transport properties of doped conjugated polymers (CPs) have been widely analyzed with the Gaussian Disorder Model (GDM) in conjunction with hopping transport between localized states. These models reveal that even in highly doped CPs, a majority of carriers are still localized because dielectric permittivity of CPs is well below that of inorganic materials, making Coulomb interactions between carriers and dopant counter-ions much more pronounced. However, previous studies within the GDM did not consider the role of screening the dielectric interactions by carriers. Here we implement carrier screening in the Debye-Hückel formalism in our calculations of dopant-induced energetic disorder, which modifies the Gaussian density of states (DOS). Then we solve the Pauli Master Equation using Miller-Abrahams hopping rates with states from the resulting screened DOS to obtain conductivity and Seebeck coefficient across a broad range of carrier concentrations and compare them to measurements. Our results show that screening has significant impact on the shape of the DOS and consequently on carrier transport, particularly at high doping. We prove that the slope of Seebeck coefficient vs electric conductivity, which was previously thought to be universal, is impacted by screening and decreases for systems with small dopant-carrier separation, explaining our measurements. 
We also show that thermoelectric power factor is underestimated by a factor of $\sim10$ at higher doping concentrations if screening is neglected. We conclude that carrier screening plays a crucial role in curtailing dopant-induced energetic disorder, particularly at high carrier concentrations.

\end{abstract}

\maketitle

Conjugated polymers (CPs) enable environmentally friendly and cost-efficient electronics from solution-processable materials. CPs are also lightweight, which makes them suitable for wearables and internet-of-things electronics \cite{Leonov2009}. They lack intrinsic free charge carriers so doping by blending small molecules with the CP emerged as an efficient way of increasing the number of carriers in CPs and improving conductivity, which is essential for applications in electronics. However, ionized dopants also act as traps. Because of their low permittivity ($\sim3$) \cite{Wang2018}, dopant counter-ions in CPs induce Coulomb interactions that alter the shape and increase the width of the intrinsic Gaussian density of states (DOS) \cite{Arkhipov2005,Upadhyaya2019,Boyle2019}, which is termed energetic disorder, hindering carrier mobility. To increase carrier concentration and optimize conductivity \cite{SalzmannACR16}, polymers are often heavily doped, with oxidation levels reaching up to 36\% \cite{BubnovaNMAT11}. 

The size of dopants and their distance from carriers determine the magnitude of Coulomb interactions and, therefore, the overall shape of the DOS. 
However, there are conflicting reports of impact of dopant features \cite{Ju2021, Murrey2022}, depending on the specific dopant-polymer system being measured \cite{Neusser2020, Li2019, Yee2019, Chen2023}. The main observables of charge transport, conductivity ($\sigma$) and Seebeck coefficient ($\alpha$), reveal information about the dynamic processes associated with carrier hopping between available sites across a range of doping concentrations. $\alpha$ is related to the average energy per carrier relative to the Fermi level and gives complementary insight into transport to $\sigma$. The interdependence of $\alpha$ and $\sigma$ plot goes beyond mere thermoelectric tradeoff as its shape provides information about energetic disorder and the DOS. Band-like transport models predict $\alpha\propto\sigma^{-1/s}$ with a slope $s=3-4$ being related to the DOS \cite{Kang2016, Lim2019, Chen2023, Choi2022, Derewjanko2022} but may require unphysically high permittivity \cite{Gregory2021} to fit experimental data. Change in the slope of this curve is often attributed to changes in the DOS \cite{Nell2018,Lim2019}, where flat slopes may indicate prominent dopant-induced energetic disorder \cite{Boyle2019}. Despite Coulomb interactions being known to dictate transport, the potential impact of carrier screening on energetic disorder and transport in CPs has not been considered before.

Using a combined computational-experimental study, we show that including carrier screening based on Debye-Hückel theory reduces dopant-induced energetic disorder and results in a simultaneous increase of $\alpha$ and $\sigma$. Our simulations, based on the Gaussian disorder model (GDM) \cite{Mendels2014} but modified to include Coulomb interactions between carriers and dopant counter-ions, solve Pauli’s master equation (PME) to simulate hopping of localized carriers from Miller-Abrahams rates \cite{Miller1960}. Our results show that screening plays a significant role in carrier mobility at high doping. Screening leads to the small slopes of $s\approx10$ in the $\alpha - \sigma$ curve observed in our measurements by limiting energetic disorder at high doping concentrations. By including dopant counter-ion size in our derivation, we were able to separate dopant radii from dopant-carrier distances, which allowed us to capture the positioning of dopant counter-ions in the host polymer matrix. We explain that dopant size has a significant impact on transport only in systems where dopants and carriers are well separated because the separation reduces the impact of carrier screening. Nevertheless, for systems with small dopant-carrier separation we observe no rapid decrease in $\alpha$, even at very high doping concentrations, and more than a ten-fold underestimation of power factor when screening is neglected.

--\textbf{Model.} To capture the impact of dopant counter-ions on carrier states, we compute the DOS after doping by adding Coulomb interactions to GDM according to the Arkhipov model \cite{Arkhipov2005}, as described in the Supplemental Information (SI) \cite{SI}. In our recent work, we found that doping leads to large energetic disorder, manifested in the form of a heavy tail in the DOS \cite{Boyle2019}, which has a dramatic impact on carrier mobility ($\mu$). Mobility in doped polymers increases with carrier concentration ($n$) as a power law $\mu\propto n^{\Gamma_E}$, where $\Gamma_E$ is the dopant-induced energetic disorder \cite{Upadhyaya2021}. Mobile carriers can rearrange in response to electric fields emanating from dopant counter-ions and partially screen the Coulomb interactions, particularly at large doping concentrations. This screening effect in CP could potentially be much more pronounced than in inorganic materials, which have larger dielectric constants. 
To capture this behavior, we base our derivation on the formalism of Debye-Hückel theory \cite{Debye1923, Kjellander1995} and the Poisson-Boltzmann equation.

\begin{figure}
	\includegraphics[width=0.47\textwidth]{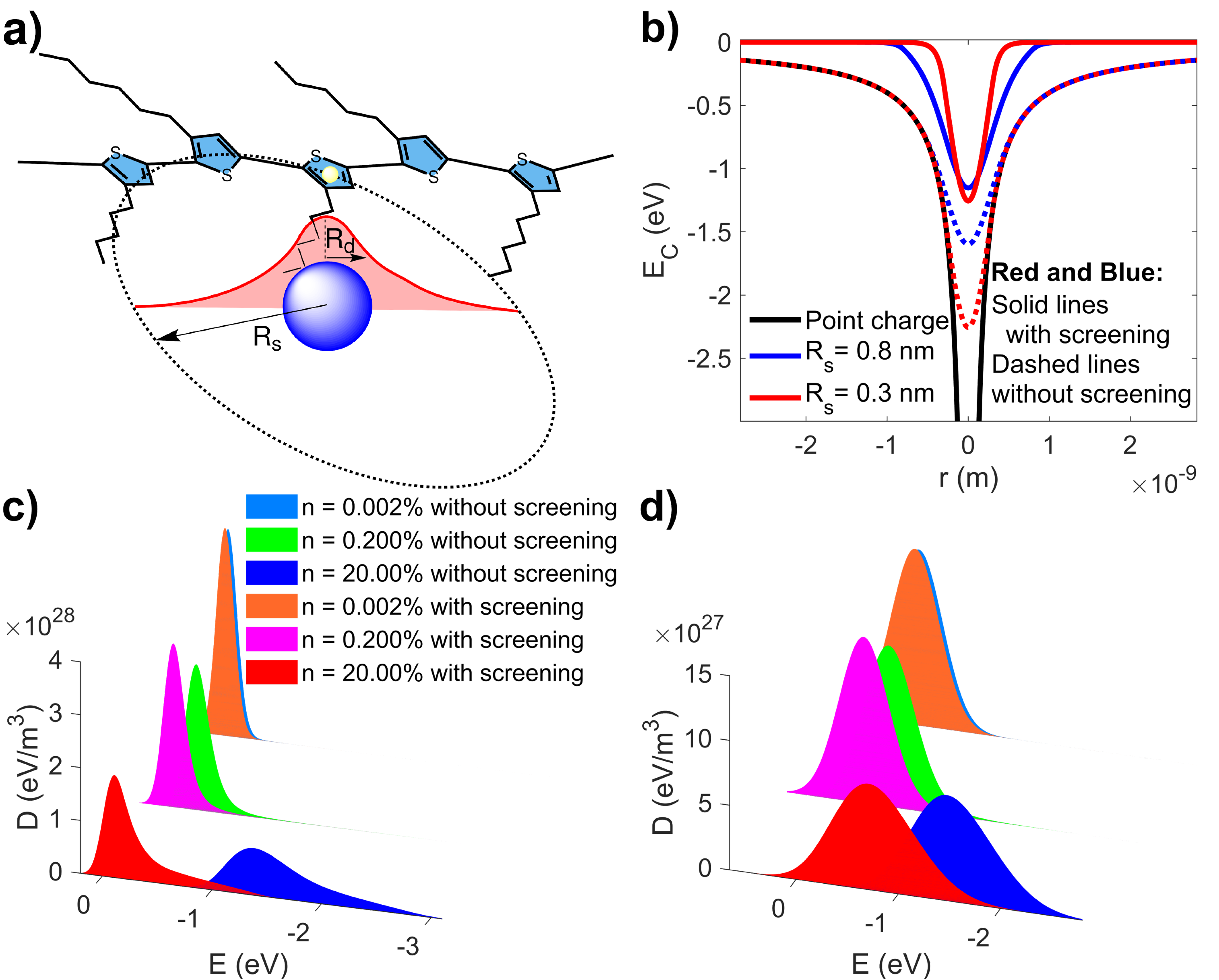}
	\caption{a) Illustration of dopant and its charge, truncated at $R_s$, with a carrier on CP's backbone. b) potential energy of the Coulomb interaction of a test charge and dopant with (solid lines) and without screening (dashed and black line). For the red lines $R_s=0.3 \text{ nm}$ and for blue $R_s=0.8 \text{ nm}$. The black solid line is the potential energy due to a point charge, simply decaying as $1/r$. Density of states with and without screening for c) $R_s=0.3 \text{ nm}$, $\Gamma_i=3 \text{ kT}$ and d) $R_s= 0.8 \text{ nm}$, $\Gamma_i=9 \text{ kT}$. The DOS tail is heavier for smaller $R_s$ and intrinsic energetic disorder $\Gamma_i$. \label{Fig1}}
\end{figure}

We expand the Poisson equation for potential \(\Phi\) into a Taylor series and retaining only the first two terms \(- \nabla^{2}\Phi = \frac{\rho(\Phi)}{\varepsilon_{0}\varepsilon} \approx \frac{\rho_{0}}{\varepsilon_{0}\varepsilon} + \frac{\Phi}{\varepsilon_{0}\varepsilon}\frac{d\rho}{d\Phi}\), where the counter-ion charge is denoted by \(\rho_{0}\) and \(\frac{d\rho}{d\Phi} = - q^{2}\frac{dn}{dE_{F}}\) accounts for the response of the carrier concentration to the potential induced by dopants. The Poisson equation then becomes
\begin{equation} \label{Eq1}
	\left( - \nabla^{2} + k_{0}^{2} \right)\Phi(r) = \frac{1}{\varepsilon_{0}\varepsilon}\rho_{0}(r),
\end{equation}
\noindent where \(k_{0} = \sqrt{\frac{1}{\varepsilon_{0}\varepsilon}q^{2}\frac{dn}{dE_{F}}}\) is the inverse screening length, $\epsilon_0$ is the permittivity of free space, and $\epsilon$ the relative permittivity of the polymer. Analytical solutions of \eqref{Eq1} have been obtained for various charge distributions (see Refs. \cite{Garavelli1991,Napsuciale2021}), including the widely used Yukawa potential for screened point charges \cite{Yukawa1955}. But in cases simultaneously involving both screening and non-trivial \(\rho_{0}\) distributions, the solution for $\Phi$ becomes quite bulky.

A Gaussian distribution has been shown to be an excellent description of the charge distribution \(\rho_{0}(r)\) for calculations of dopant-carrier interaction in the modified GDM \cite{Upadhyaya2021}. Carriers transfer from the dopant to the polymer backbone and transport along the backbone, leaving behind counter-ions that are separated from the carriers by polymer side-chains. Side-chain length, along with tight ordering of polymer chains in crystalline polymers, set the minimum distance between carriers located on the polymer backbone and dopants; this minimum separation is captured here by \(R_{s}\), illustrated in FIG. \ref{Fig1}a). Inside this radius, a carrier would be strongly trapped by the dopant and not contribute to screening so we set \(k_{0} = 0\) inside it. This distance can also not be smaller than the dopant size, so we truncate the charge distribution outside \(R_{s}\) to \(\rho_{0} = 0\).

The solution of \eqref{Eq1} then becomes \(\Phi_{out}(r) = \frac{q C_{S}}{4\pi\varepsilon_{0}\varepsilon\left( 1 + k_{0}R_{s} \right)}\frac{e^{- k_{0}(r - R_{s})}}{r}\). Here $C_S$ is the number of dopants in a cluster, here set to 1, and the term \(\left( 1 + k_{0}R_{s} \right)\) appears as a correction relative to the Yukawa potential. The solution inside \(R_{s}\) becomes:
\begin{widetext}
\begin{equation}
\label{Eq2}
\Phi_{in}(r) = \frac{q C_{S}}{4\pi\varepsilon_{0}\varepsilon R_{s}}\left\{ \frac{1}{K_{d}}\left\lbrack \frac{{erf}\left( \frac{r}{\sqrt{2}R_{d}} \right)}{r} - \frac{{erf}\left( \frac{R_{s}}{\sqrt{2}R_{d}} \right)}{R_{s}} \right\rbrack + \frac{1}{1 + k_{0}R_{s}} \right\},
\end{equation}
\end{widetext}
where \(R_{d}\) is the standard deviation of Gaussian charge distribution on the dopant counter-ion and \(K_{d} = \frac{1}{R_{s}}{erf}\left( \frac{R_{s}}{\sqrt{2}R_{d}} \right) - \sqrt{\frac{2}{\pi}}\frac{1}{R_{d}}\ \exp\left( \frac{R_{s}^{2}}{2R_{d}^2} \right)\). The resulting potential energy of a test charge interacting with a dopant counterion screened by carriers is $E_{Cin/out}(r)=-q\Phi_{in/out}(r)$, where $q$ is the elementary charge. To ensure accuracy at very large doping concentrations, we implement corrections to the screening length based on Refs. \cite{Attard1993} and \cite{Rotenberg2018}. Comparison of $k_0$ values with different corrections \cite{KruckerVelasquez2021,Kjellander1995} (FIG. S1)  and more information about our approach is in sections S2 and S3 of the SI. A large advantage of our approach is the possibility to separate \(R_{d}\), which captures the spread of the dopant counter-ion charge, and \(R_{s}\), which captures how close a carrier can approach the dopant. Thus, we are able to capture features of the dopant-polymer system more thoroughly, map those features to the model more comprehensively, and understand what features will enhance carrier transport.


--\textbf{Results.} We first examine the impact of doping on the shape of the DOS. Screening changes the electrostatic potential of dopants and consequently the DOS in the doped material. The potential energy of the Coulomb interaction of a test charge is shown in FIG. \ref{Fig1}b), contrasting the cases with (solid lines) and without (dashed lines) screening, along with the potential energy arising from unscreened Coulomb interactions with a point charge. We notice a significant change in the DOS for small \(R_{s}\) (FIG. \ref{Fig1}c)), where DOS changes shape and remains narrower. Similar effect occurs in systems with a large \(R_{s}\) (FIG. \ref{Fig1}d)), although not as pronounced due to weaker dopant-carrier interactions, resulting in a smaller impact of screening on their Coulomb potential.

\begin{figure}
\includegraphics[width=0.47\textwidth]{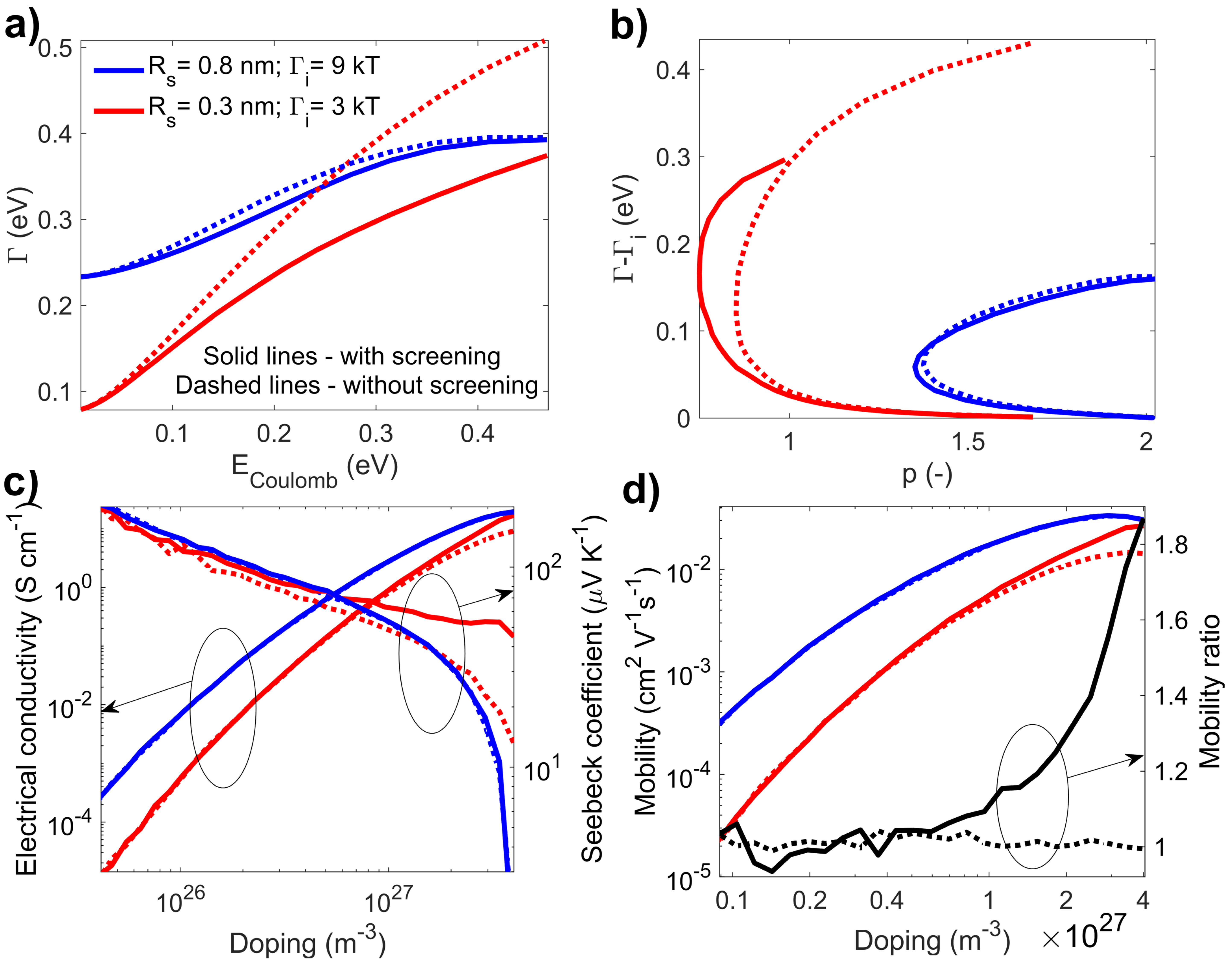}
	\caption{a) Total energetic disorder \(\Gamma\) plotted against the	Coulomb energy while varying the carrier concentration. The legend is valid for each plot. b) Extrinsic energetic disorder \(\Gamma - \Gamma_{i}\), where $\Gamma_i$ is intrinsic energetic disorder, plotted against the DOS shape parameter \(p\). Doping concentration in a) and b) ranges from $0.0002\%$ to $20\%$. Plots of c) electric conductivity and Seebeck coefficient and d) mobility as a function of doping concentration. Screening becomes significant above about $5\%$ doping where we see a significant increase in $\alpha$ in c) and mobility in d). Ratio of mobilities with and without screening is plotted on the right axis in d) for small (solid) and large (dashed) $R_s$, with \(R_{d}=0.2\) nm throughout. \label{Fig2}}
	
\end{figure}

For the purpose of analysis, we fit our DOS with a generalized Gaussian and extract the resulting energetic disorder $\Gamma$ and shape parameter $p$ from the DOS, using the procedure described in Ref. \cite{Oelerich2012}. Parameter $p$ represents DOS 'sharpness', where the DOS with \(p=2\) is Gaussian, \(p=1\) exponential, and smaller values indicate heavy tails. A quantitative example of the impact that screening has on DOS is given in FIG. \ref{Fig2}a), which shows the dependence of $\Gamma$ with Coulomb energy $E_{Coulomb}=q^2/(4\pi\varepsilon\varepsilon_0) n^{1/3}$ for screened and unscreened DOS. The density of molecular sites in our calculations is $8\cdot10^{27}\text{ m}^{-3}$ from a site spacing of $0.5\text{ nm}$. Consequently, a carrier concentration of $5\%$ corresponds to $n=4\cdot10^{26} \text{ m}^{-3}$. The evolution of the DOS with doping is shown in FIG. \ref{Fig2}b), where dopant-induced energetic disorder \(\Gamma - \Gamma_{i}\) is plotted as a function of the shape parameter \(p\). For carrier concentrations above $\sim5\%$ and small \(R_{s}\), the energetic disorder is about 0.1 eV lower in the screened than in the unscreened system. The lower energetic disorder is mirrored in the increase of $\alpha$ and $\sigma$ shown in FIG. \ref{Fig2}c). 
Consequently, carrier mobility is almost doubled at $\sim50\%$ doping when screening is included, shown in FIG. \ref{Fig2} d), where the right axis is the ratio of mobilities with and without screening, for small (solid line) and large (dashed line) $R_s$.	

The dependence of \(\sigma\) and \(\alpha\) on carrier concentration \(n\) translates into the \(\alpha - \sigma\) curve shown in FIG. 3 and elaborated in section S8 of the SI. We find that the extraordinarily flat \(\alpha - \sigma\) curves originate from screening, which is particularly pronounced at small $R_s$. In contrast, the slope of \(\alpha - \sigma\) changes only slightly with intrinsic disorder $\Gamma_i$ in the absence of screening (FIG. S6), even when comparing extreme values of $\Gamma_i$ ($3\ kT$ and $9\ kT$). We compare our calculations to experimental data on vapor-doped regiorandom (RRa) and regioregular (RR) poly(3-hexylthiophene-2,5-diyl) (P3HT). We use iodine as our dopant because its electron affinity of 4.99 eV \cite{DebesayAIMS21,Iodine} is well aligned with the HOMO level of P3HT \cite{IrwinPNAS08}, resulting in p-type material, confirmed by the positive sign of measured Seebeck coefficients. While it has been shown that energetic disorder makes dopant ionization more favorable by lifting the states in the tail of the DOS above the dopant level \cite{FediaiPCCP20,Nell2018}, we carefully confirm that we have complete dopant ionization at all doping concentrations in our model (see section S5 in the SI). Further information regarding our experimental technique can be found in our earlier work (Refs. \cite{Upadhyaya2019,Upadhyaya2021,Boyle2019}). 

\begin{figure}
	\includegraphics[width=0.44\textwidth]{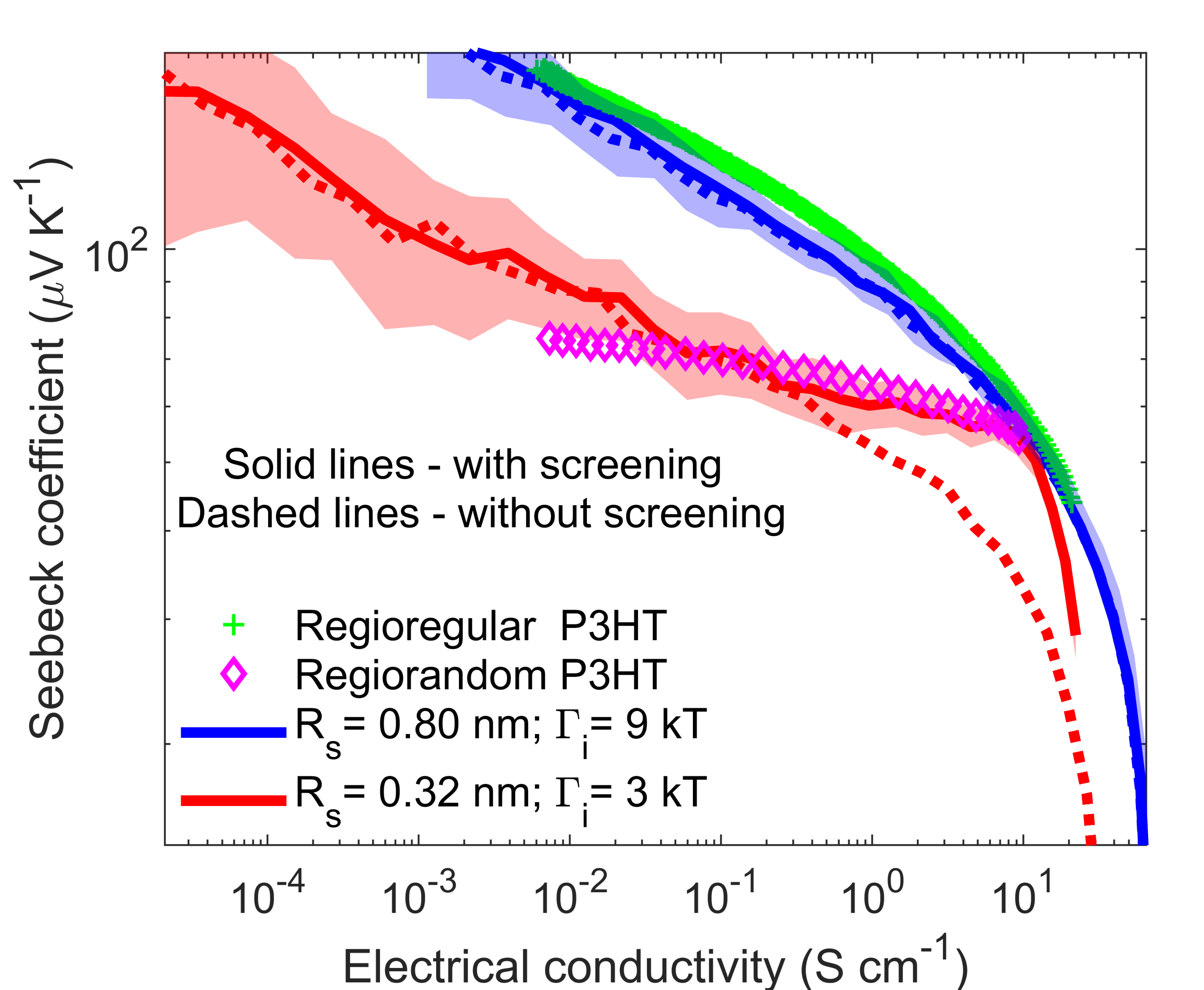}
	\caption{Seebeck coefficient against electric conductivity measurements of RR and RRa-P3HT doped with iodine and measured while dedoping. We took $R_d=0.21\text{ nm}$ to match the radius of iodide anion ($R_{I}=0.206\text{ nm}$). Dashed lines correspond to the case without screening while bands around simulation results represent variance from randomly sampling the energy sites. The attempt-to-hop frequency is 34 for RR and 10 THz for RRa P3HT. \label{Fig3}}
\end{figure}

Transport properties were continuously measured while dedoping, capturing \(\alpha - \sigma\) over several orders-of-magnitude in carrier concentration. The data is shown by symbols in FIG. \ref{Fig3} together with calculations including (solid) and excluding screening (dashed lines). Only the calculations that include screening agree with the extremely flat slope of the RRa curve. This behavior could not be explained by the unscreened GDM \cite{Upadhyaya2019} but agrees well with measurements where, even at high doping concentrations, no abrupt decrease of \(\alpha\) was observed \cite{Li2019}. It was observed previously that $10\%$ doped RR- and $20\%$ doped RRa-P3HT have similar conductivities \cite{Ukai2005}. We find that $\alpha$ of RR- and RRa-P3HT are similar at high doping due to screening, resulting in a tendency of RR and RRa-P3HT $\alpha-\sigma$ curves to intersect at high, but not identical, carrier concentrations, which our model including screening reproduces well.

\begin{figure}
    \includegraphics[width=0.47\textwidth]{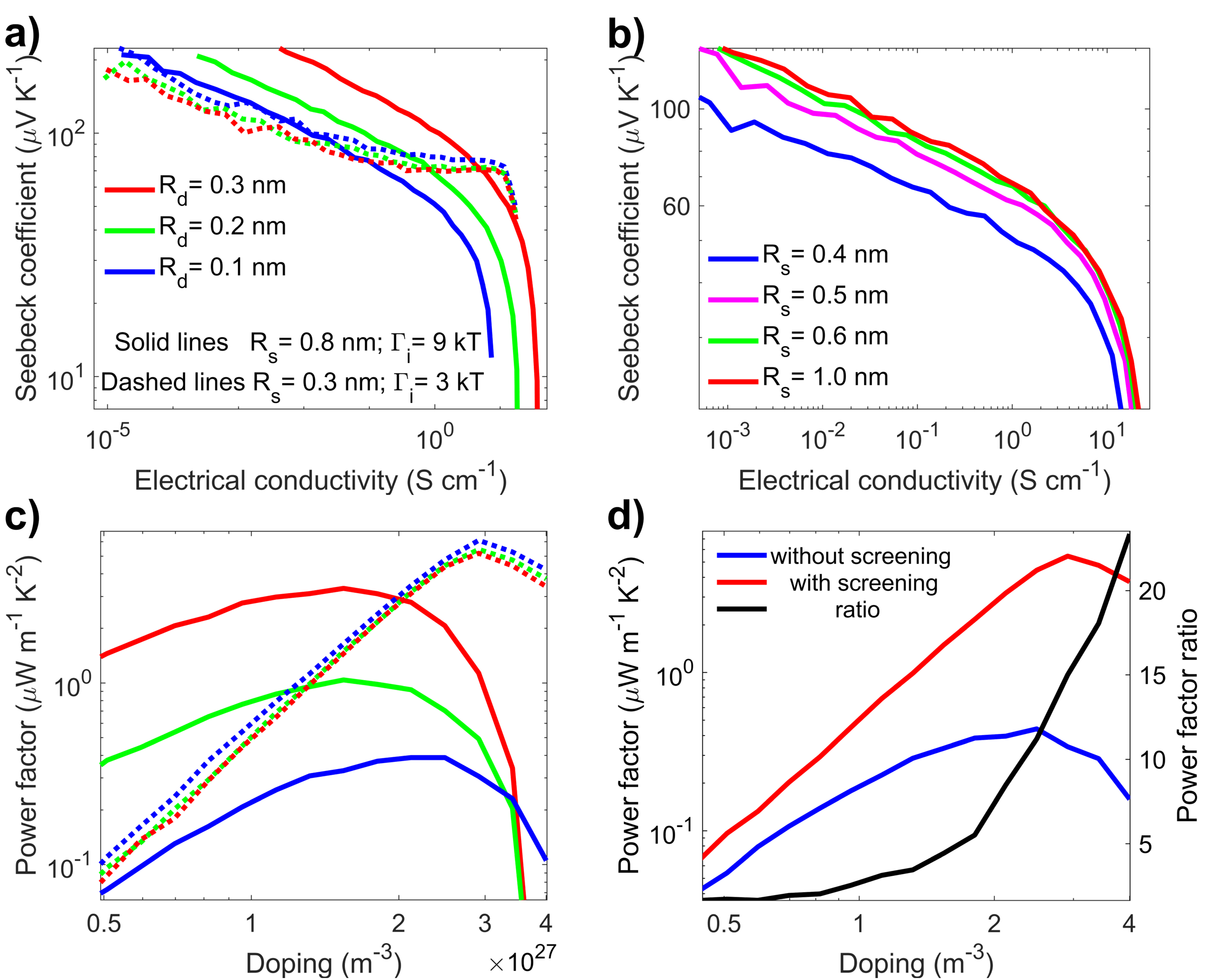}
	\caption{Seebeck coefficient plotted against electric conductivity for a Gaussian charge distribution of a) different widths \(R_{d}\) for \(R_{s} = 0.3\ nm\), \(\Gamma_{i} = 3\ kT\) and \(R_{s} = 0.8\ nm\), \(\Gamma_{i} = 9\ kT\) and b) same as in a) but with fixed \(R_{d} = 0.2\ nm\), \(\Gamma_{i} = 9\ kT\), and for varying \(R_{s}\). We observe an upward-right shift of the curve with increasing $R_d$ when $R_s$ is large, and with increasing \(R_{s}\), until saturating around \(R_{s} = 1.0\ nm\). c) Power factors vs doping with same legend as in a). Impact of screening on the power factor is shown in d) with $R_d=0.2\text{ nm}$, $R_s=0.3\text{ nm}$ and $\Gamma_i=3\text{ kT}$. The ratio of screened to unscreened case is given on the right axis, exceeding a factor of $10$ at the peak. \label{Fig4}}
\end{figure}

Measurements have also shown that dopants with larger \(R_{d}\) decrease the magnitude of Coulomb interactions, enhancing performance of CPs \cite{Ju2021}. Detrimental effect on conductivity with decreasing dopant size was also observed \cite{Jacobs2022}. This is indeed the case for large \(R_{s}\), shown in FIG. \ref{Fig4}a). When a sufficient counterion-polymer spacing is preserved, screening has less impact and increasing counterion size enhances transport properties. On the other hand, when dopants are close to the polymer backbone, screening is stronger so \(R_{d}\) has almost no effect on the \(\alpha - \sigma\) curve, shown by dashed lines in FIG. \ref{Fig4}a). This explains measurements where dopant size was observed to have no significant impact \cite{Chen2023}. Generally, increasing \(R_{s}\) shifts the \(\alpha - \sigma\) curve to the upper right, simultaneously increasing \(\sigma\) and \(\alpha\), as seen in FIG. \ref{Fig4} b), and asymptotically approaching an interaction-free limit at \(R_{s} \approx 1\ nm\), beyond which \(R_{s}\) has negligible impact. Increase in conductivity with side chain length and ordering has been observed \cite{Suh2019, Hynynen2017, HamidiSakr2017} up to a point \cite{Vijayakumar2019} where a decrease in conductivity with side chain length can be attributed to disrupted $\pi-\pi$ stacking, not considered further in our present work. Other combinations of \(R_{d}\) and \(R_{s}\) are elaborated in FIG. S5-7, section S8 of the SI. 

\begin{figure}
    \includegraphics[width=0.47\textwidth]{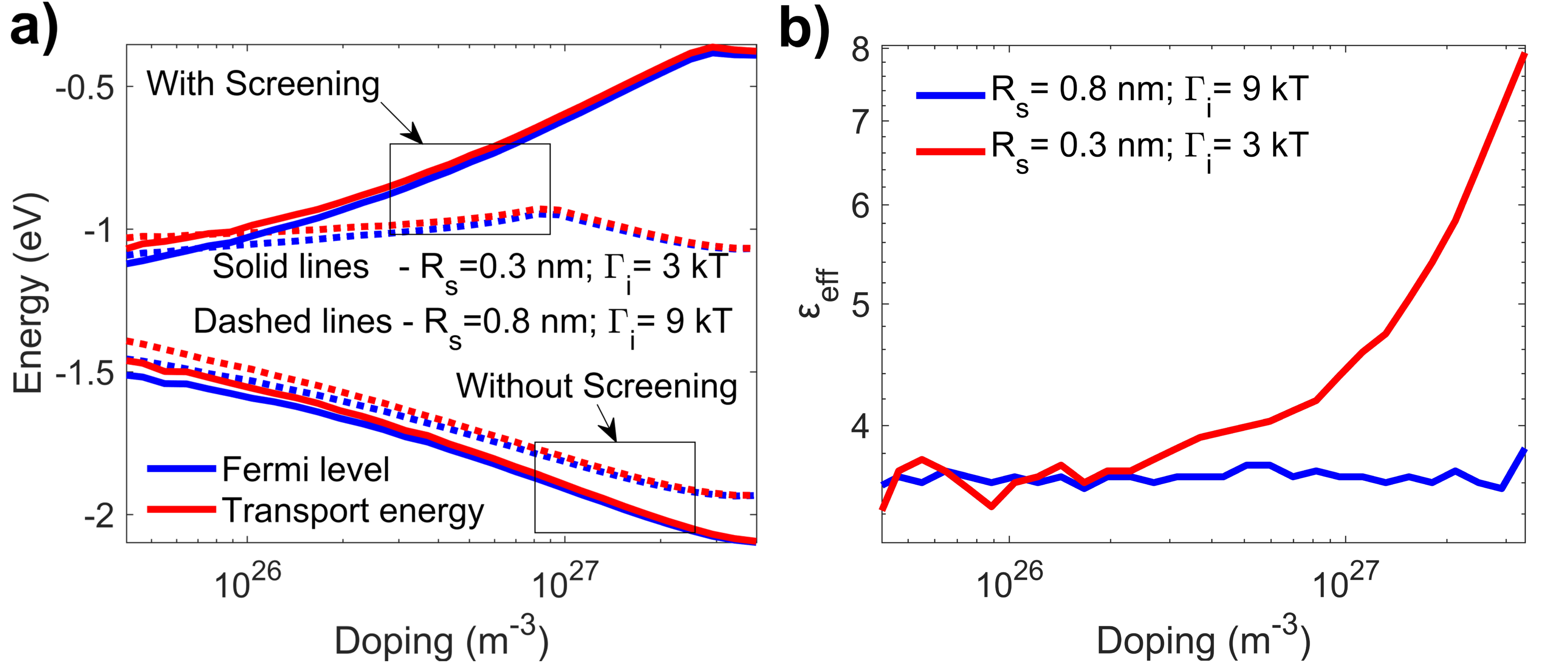}
	\caption{a) Comparison of Fermi level and transport energy between screened and unscreened system. Screening eliminates the Fermi level pinning for a system with small \(R_{s}\). b) Effective dielectric constant required to match the effect of screening as a function of doping, increasing from 3 to 8 at high doping. \label{Fig5}}
\end{figure}

Efficient thermoelectric conversion requires high power factors, defined as \(PF=\alpha^{2}\sigma\) and shown in FIG. \ref{Fig4}c). Both the magnitude of $PF$ and the doping concentration of the maximum $PF$ depend on \(R_{s}\) and \(R_{d}\). Underestimation of $PF$ without screening and for small $R_s$ highlights the impact of screening in FIG. \ref{Fig4}d), where the ratio of screened to unscreened $PF$ reaches a factor of $10$, shown on the right axis of FIG. \ref{Fig4}d).
In CPs, doping moves the Fermi level toward the center of the DOS and its position at high doping is controlled by compensation between intrinsic and dopant-induced energetic disorder \cite{Fediai2019}. Without screening, DOS shifts considerably due to the Coulomb energies added through dopants, pinning the Fermi level. For small \(R_{s}\), screening significantly reduces dopant-induced energetic disorder as well as the shift in the peak of the DOS and we observe no Fermi level pinning, as shown in FIG. \ref{Fig5}a). This implies that Fermi level and transport energy are better separated, producing higher $\alpha$.  

Here we considered interactions up to the second-nearest dopant, as further discussed in section S7 of the SI. To better illustrate the impact of screening, we extract the effective dielectric constant that the unscreened system would need to have in order to match the results with carrier screening. The procedure is described in more detail in section S6 of the SI. FIG. \ref{Fig5} d) shows effective permittivity vs carrier concentration, indicating that screening has the same effect as increasing the dielectric constant from around 3 to around 8 at high doping, in line with recent impedance spectroscopy measurements \cite{WarrenAPL23}. Analogously, metals have large concentrations of free carriers so the vicinity of metal contacts could play a role in screening Coulomb interactions, which may be a relevant strategy in organic photovoltaics \cite{KramerNanoEner22}.

--\textbf{Conclusions.} Coulomb effects have been widely established as having a significant impact on carrier transport in CPs, but the role of collective screening of these interactions by mobile carriers has not been considered. Starting from the Debye-Hückel formalism, we implement carrier screening in the modified GDM and solve the PME with states from the screened DOS. We find that screening significantly impacts transport at high doping concentrations, causing the extremely flat slopes of the \(\alpha - \sigma\) curve observed in our experiments, particularly when dopants and carriers are not well separated by the structure of the CP. 
Beyond transport, screening mitigates the shift in DOS that would lead to open-circuit voltage loss, which is of interest for photovoltaic applications \cite{Zhao2021}. 
Our work furthers the understanding of fundamental processes in CPs and may enable engineering of polymer-dopant systems for electronics and energy applications. \\

\begin{acknowledgments}
We acknowledge support from the National Science Foundation Division of Materials Research through award 2101127. We thank Dr. Scott M. Auerbach for fruitful discussions. 
\end{acknowledgments}


\begin{thebibliography}{49}%
	\makeatletter
	\providecommand \@ifxundefined [1]{%
		\@ifx{#1\undefined}
	}%
	\providecommand \@ifnum [1]{%
		\ifnum #1\expandafter \@firstoftwo
		\else \expandafter \@secondoftwo
		\fi
	}%
	\providecommand \@ifx [1]{%
		\ifx #1\expandafter \@firstoftwo
		\else \expandafter \@secondoftwo
		\fi
	}%
	\providecommand \natexlab [1]{#1}%
	\providecommand \enquote  [1]{``#1''}%
	\providecommand \bibnamefont  [1]{#1}%
	\providecommand \bibfnamefont [1]{#1}%
	\providecommand \citenamefont [1]{#1}%
	\providecommand \href@noop [0]{\@secondoftwo}%
	\providecommand \href [0]{\begingroup \@sanitize@url \@href}%
	\providecommand \@href[1]{\@@startlink{#1}\@@href}%
	\providecommand \@@href[1]{\endgroup#1\@@endlink}%
	\providecommand \@sanitize@url [0]{\catcode `\\12\catcode `\$12\catcode
		`\&12\catcode `\#12\catcode `\^12\catcode `\_12\catcode `\%12\relax}%
	\providecommand \@@startlink[1]{}%
	\providecommand \@@endlink[0]{}%
	\providecommand \url  [0]{\begingroup\@sanitize@url \@url }%
	\providecommand \@url [1]{\endgroup\@href {#1}{\urlprefix }}%
	\providecommand \urlprefix  [0]{URL }%
	\providecommand \Eprint [0]{\href }%
	\providecommand \doibase [0]{https://doi.org/}%
	\providecommand \selectlanguage [0]{\@gobble}%
	\providecommand \bibinfo  [0]{\@secondoftwo}%
	\providecommand \bibfield  [0]{\@secondoftwo}%
	\providecommand \translation [1]{[#1]}%
	\providecommand \BibitemOpen [0]{}%
	\providecommand \bibitemStop [0]{}%
	\providecommand \bibitemNoStop [0]{.\EOS\space}%
	\providecommand \EOS [0]{\spacefactor3000\relax}%
	\providecommand \BibitemShut  [1]{\csname bibitem#1\endcsname}%
	\let\auto@bib@innerbib\@empty
	\bibitem [{\citenamefont {Leonov}\ and\ \citenamefont
		{Vullers}(2009)}]{Leonov2009}%
	\BibitemOpen
	\bibfield  {author} {\bibinfo {author} {\bibfnamefont {V.}~\bibnamefont
			{Leonov}}\ and\ \bibinfo {author} {\bibfnamefont {R.~J.~M.}\ \bibnamefont
			{Vullers}},\ }\href@noop {} {\bibfield  {journal} {\bibinfo  {journal}
			{Journal of Renewable and Sustainable Energy}\ }\textbf {\bibinfo {volume}
			{1}},\ \bibinfo {pages} {062701} (\bibinfo {year} {2009})}\BibitemShut
	{NoStop}%
	\bibitem [{\citenamefont {Wang}\ \emph {et~al.}(2018)\citenamefont {Wang},
		\citenamefont {Zhang}, \citenamefont {Peji{\'{c}}}, \citenamefont {Li},
		\citenamefont {Fukuto}, \citenamefont {Zhu},\ and\ \citenamefont
		{Sauv{\'{e}}}}]{Wang2018}%
	\BibitemOpen
	\bibfield  {author} {\bibinfo {author} {\bibfnamefont {C.}~\bibnamefont
			{Wang}}, \bibinfo {author} {\bibfnamefont {Z.}~\bibnamefont {Zhang}},
		\bibinfo {author} {\bibfnamefont {S.}~\bibnamefont {Peji{\'{c}}}}, \bibinfo
		{author} {\bibfnamefont {R.}~\bibnamefont {Li}}, \bibinfo {author}
		{\bibfnamefont {M.}~\bibnamefont {Fukuto}}, \bibinfo {author} {\bibfnamefont
			{L.}~\bibnamefont {Zhu}},\ and\ \bibinfo {author} {\bibfnamefont
			{G.}~\bibnamefont {Sauv{\'{e}}}},\ }\href@noop {} {\bibfield  {journal}
		{\bibinfo  {journal} {Macromolecules}\ }\textbf {\bibinfo {volume} {51}},\
		\bibinfo {pages} {9368} (\bibinfo {year} {2018})}\BibitemShut {NoStop}%
	\bibitem [{\citenamefont {Arkhipov}\ \emph {et~al.}(2005)\citenamefont
		{Arkhipov}, \citenamefont {Heremans}, \citenamefont {Emelianova},\ and\
		\citenamefont {Bässler}}]{Arkhipov2005}%
	\BibitemOpen
	\bibfield  {author} {\bibinfo {author} {\bibfnamefont {V.~I.}\ \bibnamefont
			{Arkhipov}}, \bibinfo {author} {\bibfnamefont {P.}~\bibnamefont {Heremans}},
		\bibinfo {author} {\bibfnamefont {E.~V.}\ \bibnamefont {Emelianova}},\ and\
		\bibinfo {author} {\bibfnamefont {H.}~\bibnamefont {Bässler}},\ }\href@noop
	{} {\bibfield  {journal} {\bibinfo  {journal} {Phys. Rev. B}\ }\textbf
		{\bibinfo {volume} {71}},\ \bibinfo {pages} {045214} (\bibinfo {year}
		{2005})}\BibitemShut {NoStop}%
	\bibitem [{\citenamefont {Upadhyaya}\ \emph {et~al.}(2019)\citenamefont
		{Upadhyaya}, \citenamefont {Boyle}, \citenamefont {Venkataraman},\ and\
		\citenamefont {Aksamija}}]{Upadhyaya2019}%
	\BibitemOpen
	\bibfield  {author} {\bibinfo {author} {\bibfnamefont {M.}~\bibnamefont
			{Upadhyaya}}, \bibinfo {author} {\bibfnamefont {C.~J.}\ \bibnamefont
			{Boyle}}, \bibinfo {author} {\bibfnamefont {D.}~\bibnamefont
			{Venkataraman}},\ and\ \bibinfo {author} {\bibfnamefont {Z.}~\bibnamefont
			{Aksamija}},\ }\href@noop {} {\bibfield  {journal} {\bibinfo  {journal} {Sci.
				Rep.}\ }\textbf {\bibinfo {volume} {9}} (\bibinfo {year} {2019})}\BibitemShut
	{NoStop}%
	\bibitem [{\citenamefont {Boyle}\ \emph {et~al.}(2019)\citenamefont {Boyle},
		\citenamefont {Upadhyaya}, \citenamefont {Wang}, \citenamefont {Renna},
		\citenamefont {Lu-D{\'{\i}}az}, \citenamefont {Jeong}, \citenamefont
		{Hight-Huf}, \citenamefont {Korugic-Karasz}, \citenamefont {Barnes},
		\citenamefont {Aksamija},\ and\ \citenamefont {Venkataraman}}]{Boyle2019}%
	\BibitemOpen
	\bibfield  {author} {\bibinfo {author} {\bibfnamefont {C.~J.}\ \bibnamefont
			{Boyle}}, \bibinfo {author} {\bibfnamefont {M.}~\bibnamefont {Upadhyaya}},
		\bibinfo {author} {\bibfnamefont {P.}~\bibnamefont {Wang}}, \bibinfo {author}
		{\bibfnamefont {L.~A.}\ \bibnamefont {Renna}}, \bibinfo {author}
		{\bibfnamefont {M.}~\bibnamefont {Lu-D{\'{\i}}az}}, \bibinfo {author}
		{\bibfnamefont {S.~P.}\ \bibnamefont {Jeong}}, \bibinfo {author}
		{\bibfnamefont {N.}~\bibnamefont {Hight-Huf}}, \bibinfo {author}
		{\bibfnamefont {L.}~\bibnamefont {Korugic-Karasz}}, \bibinfo {author}
		{\bibfnamefont {M.~D.}\ \bibnamefont {Barnes}}, \bibinfo {author}
		{\bibfnamefont {Z.}~\bibnamefont {Aksamija}},\ and\ \bibinfo {author}
		{\bibfnamefont {D.}~\bibnamefont {Venkataraman}},\ }\href@noop {} {\bibfield
		{journal} {\bibinfo  {journal} {Nat. Commun.}\ }\textbf {\bibinfo {volume}
			{10}} (\bibinfo {year} {2019})}\BibitemShut {NoStop}%
	\bibitem [{\citenamefont {Salzmann}\ \emph {et~al.}(2016)\citenamefont
		{Salzmann}, \citenamefont {Heimel}, \citenamefont {Oehzelt}, \citenamefont
		{Winkler},\ and\ \citenamefont {Koch}}]{SalzmannACR16}%
	\BibitemOpen
	\bibfield  {author} {\bibinfo {author} {\bibfnamefont {I.}~\bibnamefont
			{Salzmann}}, \bibinfo {author} {\bibfnamefont {G.}~\bibnamefont {Heimel}},
		\bibinfo {author} {\bibfnamefont {M.}~\bibnamefont {Oehzelt}}, \bibinfo
		{author} {\bibfnamefont {S.}~\bibnamefont {Winkler}},\ and\ \bibinfo {author}
		{\bibfnamefont {N.}~\bibnamefont {Koch}},\ }\href@noop {} {\bibfield
		{journal} {\bibinfo  {journal} {Accounts Chem. Res.}\ }\textbf {\bibinfo
			{volume} {49}},\ \bibinfo {pages} {370} (\bibinfo {year} {2016})}\BibitemShut
	{NoStop}%
	\bibitem [{\citenamefont {Bubnova}\ \emph {et~al.}(2011)\citenamefont
		{Bubnova}, \citenamefont {Khan}, \citenamefont {Malti}, \citenamefont
		{Braun}, \citenamefont {Fahlman}, \citenamefont {Berggren},\ and\
		\citenamefont {Crispin}}]{BubnovaNMAT11}%
	\BibitemOpen
	\bibfield  {author} {\bibinfo {author} {\bibfnamefont {O.}~\bibnamefont
			{Bubnova}}, \bibinfo {author} {\bibfnamefont {Z.~U.}\ \bibnamefont {Khan}},
		\bibinfo {author} {\bibfnamefont {A.}~\bibnamefont {Malti}}, \bibinfo
		{author} {\bibfnamefont {S.}~\bibnamefont {Braun}}, \bibinfo {author}
		{\bibfnamefont {M.}~\bibnamefont {Fahlman}}, \bibinfo {author} {\bibfnamefont
			{M.}~\bibnamefont {Berggren}},\ and\ \bibinfo {author} {\bibfnamefont
			{X.}~\bibnamefont {Crispin}},\ }\href@noop {} {\bibfield  {journal} {\bibinfo
			{journal} {Nature Mater.}\ }\textbf {\bibinfo {volume} {10}},\ \bibinfo
		{pages} {429} (\bibinfo {year} {2011})}\BibitemShut {NoStop}%
	\bibitem [{\citenamefont {Ju}\ \emph {et~al.}(2021)\citenamefont {Ju},
		\citenamefont {Kim}, \citenamefont {Yook}, \citenamefont {Han},\ and\
		\citenamefont {Cho}}]{Ju2021}%
	\BibitemOpen
	\bibfield  {author} {\bibinfo {author} {\bibfnamefont {D.}~\bibnamefont
			{Ju}}, \bibinfo {author} {\bibfnamefont {J.}~\bibnamefont {Kim}}, \bibinfo
		{author} {\bibfnamefont {H.}~\bibnamefont {Yook}}, \bibinfo {author}
		{\bibfnamefont {J.~W.}\ \bibnamefont {Han}},\ and\ \bibinfo {author}
		{\bibfnamefont {K.}~\bibnamefont {Cho}},\ }\href@noop {} {\bibfield
		{journal} {\bibinfo  {journal} {Nano Energy}\ }\textbf {\bibinfo {volume}
			{90}},\ \bibinfo {pages} {106604} (\bibinfo {year} {2021})}\BibitemShut
	{NoStop}%
	\bibitem [{\citenamefont {Murrey}\ \emph {et~al.}(2022)\citenamefont {Murrey},
		\citenamefont {Aubry}, \citenamefont {Ruiz}, \citenamefont {Thurman},
		\citenamefont {Eckstein}, \citenamefont {Doud}, \citenamefont {Stauber},
		\citenamefont {Spokoyny}, \citenamefont {Schwartz}, \citenamefont {Hertel},
		\citenamefont {Blackburn},\ and\ \citenamefont {Ferguson}}]{Murrey2022}%
	\BibitemOpen
	\bibfield  {author} {\bibinfo {author} {\bibfnamefont {T.~L.}\ \bibnamefont
			{Murrey}}, \bibinfo {author} {\bibfnamefont {T.~J.}\ \bibnamefont {Aubry}},
		\bibinfo {author} {\bibfnamefont {O.~L.}\ \bibnamefont {Ruiz}}, \bibinfo
		{author} {\bibfnamefont {K.~A.}\ \bibnamefont {Thurman}}, \bibinfo {author}
		{\bibfnamefont {K.~H.}\ \bibnamefont {Eckstein}}, \bibinfo {author}
		{\bibfnamefont {E.~A.}\ \bibnamefont {Doud}}, \bibinfo {author}
		{\bibfnamefont {J.~M.}\ \bibnamefont {Stauber}}, \bibinfo {author}
		{\bibfnamefont {A.~M.}\ \bibnamefont {Spokoyny}}, \bibinfo {author}
		{\bibfnamefont {B.~J.}\ \bibnamefont {Schwartz}}, \bibinfo {author}
		{\bibfnamefont {T.}~\bibnamefont {Hertel}}, \bibinfo {author} {\bibfnamefont
			{J.}~\bibnamefont {Blackburn}},\ and\ \bibinfo {author} {\bibfnamefont
			{A.~J.}\ \bibnamefont {Ferguson}},\ }\href@noop {} {\bibfield  {journal}
		{\bibinfo  {journal} {{SSRN} Electronic Journal}\ } (\bibinfo {year}
		{2022})}\BibitemShut {NoStop}%
	\bibitem [{\citenamefont {Neusser}\ \emph {et~al.}(2020)\citenamefont
		{Neusser}, \citenamefont {Malacrida}, \citenamefont {Kern}, \citenamefont
		{Gross}, \citenamefont {van Slageren},\ and\ \citenamefont
		{Ludwigs}}]{Neusser2020}%
	\BibitemOpen
	\bibfield  {author} {\bibinfo {author} {\bibfnamefont {D.}~\bibnamefont
			{Neusser}}, \bibinfo {author} {\bibfnamefont {C.}~\bibnamefont {Malacrida}},
		\bibinfo {author} {\bibfnamefont {M.}~\bibnamefont {Kern}}, \bibinfo {author}
		{\bibfnamefont {Y.~M.}\ \bibnamefont {Gross}}, \bibinfo {author}
		{\bibfnamefont {J.}~\bibnamefont {van Slageren}},\ and\ \bibinfo {author}
		{\bibfnamefont {S.}~\bibnamefont {Ludwigs}},\ }\href@noop {} {\bibfield
		{journal} {\bibinfo  {journal} {Chem. Mater.}\ }\textbf {\bibinfo {volume}
			{32}},\ \bibinfo {pages} {6003} (\bibinfo {year} {2020})}\BibitemShut
	{NoStop}%
	\bibitem [{\citenamefont {Li}\ \emph {et~al.}(2019)\citenamefont {Li},
		\citenamefont {Plunkett}, \citenamefont {Cai}, \citenamefont {Qiu},
		\citenamefont {Wei}, \citenamefont {Chen}, \citenamefont {Thon},
		\citenamefont {Reich}, \citenamefont {Chen},\ and\ \citenamefont
		{Katz}}]{Li2019}%
	\BibitemOpen
	\bibfield  {author} {\bibinfo {author} {\bibfnamefont {H.}~\bibnamefont
			{Li}}, \bibinfo {author} {\bibfnamefont {E.}~\bibnamefont {Plunkett}},
		\bibinfo {author} {\bibfnamefont {Z.}~\bibnamefont {Cai}}, \bibinfo {author}
		{\bibfnamefont {B.}~\bibnamefont {Qiu}}, \bibinfo {author} {\bibfnamefont
			{T.}~\bibnamefont {Wei}}, \bibinfo {author} {\bibfnamefont {H.}~\bibnamefont
			{Chen}}, \bibinfo {author} {\bibfnamefont {S.~M.}\ \bibnamefont {Thon}},
		\bibinfo {author} {\bibfnamefont {D.~H.}\ \bibnamefont {Reich}}, \bibinfo
		{author} {\bibfnamefont {L.}~\bibnamefont {Chen}},\ and\ \bibinfo {author}
		{\bibfnamefont {H.~E.}\ \bibnamefont {Katz}},\ }\href@noop {} {\bibfield
		{journal} {\bibinfo  {journal} {Adv. Electron. Mater.}\ }\textbf {\bibinfo
			{volume} {5}},\ \bibinfo {pages} {1800618} (\bibinfo {year}
		{2019})}\BibitemShut {NoStop}%
	\bibitem [{\citenamefont {Yee}\ \emph {et~al.}(2019)\citenamefont {Yee},
		\citenamefont {Scholes}, \citenamefont {Schwartz},\ and\ \citenamefont
		{Tolbert}}]{Yee2019}%
	\BibitemOpen
	\bibfield  {author} {\bibinfo {author} {\bibfnamefont {P.~Y.}\ \bibnamefont
			{Yee}}, \bibinfo {author} {\bibfnamefont {D.~T.}\ \bibnamefont {Scholes}},
		\bibinfo {author} {\bibfnamefont {B.~J.}\ \bibnamefont {Schwartz}},\ and\
		\bibinfo {author} {\bibfnamefont {S.~H.}\ \bibnamefont {Tolbert}},\
	}\href@noop {} {\bibfield  {journal} {\bibinfo  {journal} {The Journal of
				Physical Chemistry Letters}\ }\textbf {\bibinfo {volume} {10}},\ \bibinfo
		{pages} {4929} (\bibinfo {year} {2019})}\BibitemShut {NoStop}%
	\bibitem [{\citenamefont {Chen}\ \emph {et~al.}(2023)\citenamefont {Chen},
		\citenamefont {Jacobs}, \citenamefont {Kang}, \citenamefont {Lin},
		\citenamefont {Jellett}, \citenamefont {Kang}, \citenamefont {Lee},
		\citenamefont {Huang}, \citenamefont {BaloochQarai}, \citenamefont {Ghosh},
		\citenamefont {Statz}, \citenamefont {Wood}, \citenamefont {Ren},
		\citenamefont {Tjhe}, \citenamefont {Sun}, \citenamefont {She}, \citenamefont
		{Hu}, \citenamefont {Jiang}, \citenamefont {Spano}, \citenamefont
		{McCulloch},\ and\ \citenamefont {Sirringhaus}}]{Chen2023}%
	\BibitemOpen
	\bibfield  {author} {\bibinfo {author} {\bibfnamefont {C.}~\bibnamefont
			{Chen}}, \bibinfo {author} {\bibfnamefont {I.~E.}\ \bibnamefont {Jacobs}},
		\bibinfo {author} {\bibfnamefont {K.}~\bibnamefont {Kang}}, \bibinfo {author}
		{\bibfnamefont {Y.}~\bibnamefont {Lin}}, \bibinfo {author} {\bibfnamefont
			{C.}~\bibnamefont {Jellett}}, \bibinfo {author} {\bibfnamefont
			{B.}~\bibnamefont {Kang}}, \bibinfo {author} {\bibfnamefont {S.~B.}\
			\bibnamefont {Lee}}, \bibinfo {author} {\bibfnamefont {Y.}~\bibnamefont
			{Huang}}, \bibinfo {author} {\bibfnamefont {M.}~\bibnamefont {BaloochQarai}},
		\bibinfo {author} {\bibfnamefont {R.}~\bibnamefont {Ghosh}}, \bibinfo
		{author} {\bibfnamefont {M.}~\bibnamefont {Statz}}, \bibinfo {author}
		{\bibfnamefont {W.}~\bibnamefont {Wood}}, \bibinfo {author} {\bibfnamefont
			{X.}~\bibnamefont {Ren}}, \bibinfo {author} {\bibfnamefont {D.}~\bibnamefont
			{Tjhe}}, \bibinfo {author} {\bibfnamefont {Y.}~\bibnamefont {Sun}}, \bibinfo
		{author} {\bibfnamefont {X.}~\bibnamefont {She}}, \bibinfo {author}
		{\bibfnamefont {Y.}~\bibnamefont {Hu}}, \bibinfo {author} {\bibfnamefont
			{L.}~\bibnamefont {Jiang}}, \bibinfo {author} {\bibfnamefont {F.~C.}\
			\bibnamefont {Spano}}, \bibinfo {author} {\bibfnamefont {I.}~\bibnamefont
			{McCulloch}},\ and\ \bibinfo {author} {\bibfnamefont {H.}~\bibnamefont
			{Sirringhaus}},\ }\href@noop {} {\bibfield  {journal} {\bibinfo  {journal}
			{Adv. Energy Mater.}\ ,\ \bibinfo {pages} {2202797}} (\bibinfo {year}
		{2023})}\BibitemShut {NoStop}%
	\bibitem [{\citenamefont {Kang}\ and\ \citenamefont {Snyder}(2016)}]{Kang2016}%
	\BibitemOpen
	\bibfield  {author} {\bibinfo {author} {\bibfnamefont {S.~D.}\ \bibnamefont
			{Kang}}\ and\ \bibinfo {author} {\bibfnamefont {G.~J.}\ \bibnamefont
			{Snyder}},\ }\href@noop {} {\bibfield  {journal} {\bibinfo  {journal} {Nature
				Mater.}\ }\textbf {\bibinfo {volume} {16}},\ \bibinfo {pages} {252} (\bibinfo
		{year} {2016})}\BibitemShut {NoStop}%
	\bibitem [{\citenamefont {Lim}\ \emph {et~al.}(2019)\citenamefont {Lim},
		\citenamefont {Glaudell}, \citenamefont {Miller},\ and\ \citenamefont
		{Chabinyc}}]{Lim2019}%
	\BibitemOpen
	\bibfield  {author} {\bibinfo {author} {\bibfnamefont {E.}~\bibnamefont
			{Lim}}, \bibinfo {author} {\bibfnamefont {A.~M.}\ \bibnamefont {Glaudell}},
		\bibinfo {author} {\bibfnamefont {R.}~\bibnamefont {Miller}},\ and\ \bibinfo
		{author} {\bibfnamefont {M.~L.}\ \bibnamefont {Chabinyc}},\ }\href@noop {}
	{\bibfield  {journal} {\bibinfo  {journal} {Adv. Electron. Mater.}\ }\textbf
		{\bibinfo {volume} {5}},\ \bibinfo {pages} {1800915} (\bibinfo {year}
		{2019})}\BibitemShut {NoStop}%
	\bibitem [{\citenamefont {Choi}\ \emph {et~al.}(2022)\citenamefont {Choi},
		\citenamefont {Kim}, \citenamefont {Lee}, \citenamefont {Jung}, \citenamefont
		{Tripathi}, \citenamefont {Lee}, \citenamefont {Woo},\ and\ \citenamefont
		{Lee}}]{Choi2022}%
	\BibitemOpen
	\bibfield  {author} {\bibinfo {author} {\bibfnamefont {W.}~\bibnamefont
			{Choi}}, \bibinfo {author} {\bibfnamefont {S.}~\bibnamefont {Kim}}, \bibinfo
		{author} {\bibfnamefont {S.}~\bibnamefont {Lee}}, \bibinfo {author}
		{\bibfnamefont {C.}~\bibnamefont {Jung}}, \bibinfo {author} {\bibfnamefont
			{A.}~\bibnamefont {Tripathi}}, \bibinfo {author} {\bibfnamefont
			{Y.}~\bibnamefont {Lee}}, \bibinfo {author} {\bibfnamefont {H.~Y.}\
			\bibnamefont {Woo}},\ and\ \bibinfo {author} {\bibfnamefont {H.}~\bibnamefont
			{Lee}},\ }\href@noop {} {\bibfield  {journal} {\bibinfo  {journal} {Small
				Methods}\ ,\ \bibinfo {pages} {2201145}} (\bibinfo {year}
		{2022})}\BibitemShut {NoStop}%
	\bibitem [{\citenamefont {Derewjanko}\ \emph {et~al.}(2022)\citenamefont
		{Derewjanko}, \citenamefont {Scheunemann}, \citenamefont {Järsvall},
		\citenamefont {Hofmann}, \citenamefont {Müller},\ and\ \citenamefont
		{Kemerink}}]{Derewjanko2022}%
	\BibitemOpen
	\bibfield  {author} {\bibinfo {author} {\bibfnamefont {D.}~\bibnamefont
			{Derewjanko}}, \bibinfo {author} {\bibfnamefont {D.}~\bibnamefont
			{Scheunemann}}, \bibinfo {author} {\bibfnamefont {E.}~\bibnamefont
			{Järsvall}}, \bibinfo {author} {\bibfnamefont {A.~I.}\ \bibnamefont
			{Hofmann}}, \bibinfo {author} {\bibfnamefont {C.}~\bibnamefont {Müller}},\
		and\ \bibinfo {author} {\bibfnamefont {M.}~\bibnamefont {Kemerink}},\
	}\href@noop {} {\bibfield  {journal} {\bibinfo  {journal} {Adv. Funct.
				Mater.}\ }\textbf {\bibinfo {volume} {32}},\ \bibinfo {pages} {2112262}
		(\bibinfo {year} {2022})}\BibitemShut {NoStop}%
	\bibitem [{\citenamefont {Gregory}\ \emph {et~al.}(2021)\citenamefont
		{Gregory}, \citenamefont {Hanus}, \citenamefont {Atassi}, \citenamefont
		{Rinehart}, \citenamefont {Wooding}, \citenamefont {Menon}, \citenamefont
		{Losego}, \citenamefont {Snyder},\ and\ \citenamefont {Yee}}]{Gregory2021}%
	\BibitemOpen
	\bibfield  {author} {\bibinfo {author} {\bibfnamefont {S.~A.}\ \bibnamefont
			{Gregory}}, \bibinfo {author} {\bibfnamefont {R.}~\bibnamefont {Hanus}},
		\bibinfo {author} {\bibfnamefont {A.}~\bibnamefont {Atassi}}, \bibinfo
		{author} {\bibfnamefont {J.~M.}\ \bibnamefont {Rinehart}}, \bibinfo {author}
		{\bibfnamefont {J.~P.}\ \bibnamefont {Wooding}}, \bibinfo {author}
		{\bibfnamefont {A.~K.}\ \bibnamefont {Menon}}, \bibinfo {author}
		{\bibfnamefont {M.~D.}\ \bibnamefont {Losego}}, \bibinfo {author}
		{\bibfnamefont {G.~J.}\ \bibnamefont {Snyder}},\ and\ \bibinfo {author}
		{\bibfnamefont {S.~K.}\ \bibnamefont {Yee}},\ }\href@noop {} {\bibfield
		{journal} {\bibinfo  {journal} {Nat. Mater.}\ }\textbf {\bibinfo {volume}
			{20}},\ \bibinfo {pages} {1414} (\bibinfo {year} {2021})}\BibitemShut
	{NoStop}%
	\bibitem [{\citenamefont {Nell}\ \emph {et~al.}(2018)\citenamefont {Nell},
		\citenamefont {Ortstein}, \citenamefont {Boltalina},\ and\ \citenamefont
		{Vandewal}}]{Nell2018}%
	\BibitemOpen
	\bibfield  {author} {\bibinfo {author} {\bibfnamefont {B.}~\bibnamefont
			{Nell}}, \bibinfo {author} {\bibfnamefont {K.}~\bibnamefont {Ortstein}},
		\bibinfo {author} {\bibfnamefont {O.~V.}\ \bibnamefont {Boltalina}},\ and\
		\bibinfo {author} {\bibfnamefont {K.}~\bibnamefont {Vandewal}},\ }\href@noop
	{} {\bibfield  {journal} {\bibinfo  {journal} {J. Phys. Chem. C}\ }\textbf
		{\bibinfo {volume} {122}},\ \bibinfo {pages} {11730} (\bibinfo {year}
		{2018})}\BibitemShut {NoStop}%
	\bibitem [{\citenamefont {Mendels}\ and\ \citenamefont
		{Tessler}(2014)}]{Mendels2014}%
	\BibitemOpen
	\bibfield  {author} {\bibinfo {author} {\bibfnamefont {D.}~\bibnamefont
			{Mendels}}\ and\ \bibinfo {author} {\bibfnamefont {N.}~\bibnamefont
			{Tessler}},\ }\href {https://doi.org/10.1021/jz5016058} {\bibfield  {journal}
		{\bibinfo  {journal} {The Journal of Physical Chemistry Letters}\ }\textbf
		{\bibinfo {volume} {5}},\ \bibinfo {pages} {3247} (\bibinfo {year}
		{2014})}\BibitemShut {NoStop}%
	\bibitem [{\citenamefont {Miller}\ and\ \citenamefont
		{Abrahams}(1960)}]{Miller1960}%
	\BibitemOpen
	\bibfield  {author} {\bibinfo {author} {\bibfnamefont {A.}~\bibnamefont
			{Miller}}\ and\ \bibinfo {author} {\bibfnamefont {E.}~\bibnamefont
			{Abrahams}},\ }\href {https://doi.org/10.1103/physrev.120.745} {\bibfield
		{journal} {\bibinfo  {journal} {Physical Review}\ }\textbf {\bibinfo {volume}
			{120}},\ \bibinfo {pages} {745} (\bibinfo {year} {1960})}\BibitemShut
	{NoStop}%
	\bibitem [{SI()}]{SI}%
	\BibitemOpen
	\href@noop {} {\emph {\bibinfo {title} {See Supplemental Material [URL] for
				more details on DOS tail broadening, which includes Refs.
				\cite{Zuo2016,Zuo2018}}}}\BibitemShut {NoStop}%
	\bibitem [{\citenamefont {Upadhyaya}\ \emph {et~al.}(2021)\citenamefont
		{Upadhyaya}, \citenamefont {Lu-D{\'{\i}}az}, \citenamefont {Samanta},
		\citenamefont {Abdullah}, \citenamefont {Dusoe}, \citenamefont {Kittilstved},
		\citenamefont {Venkataraman},\ and\ \citenamefont
		{Ak{\v{s}}amija}}]{Upadhyaya2021}%
	\BibitemOpen
	\bibfield  {author} {\bibinfo {author} {\bibfnamefont {M.}~\bibnamefont
			{Upadhyaya}}, \bibinfo {author} {\bibfnamefont {M.}~\bibnamefont
			{Lu-D{\'{\i}}az}}, \bibinfo {author} {\bibfnamefont {S.}~\bibnamefont
			{Samanta}}, \bibinfo {author} {\bibfnamefont {M.}~\bibnamefont {Abdullah}},
		\bibinfo {author} {\bibfnamefont {K.}~\bibnamefont {Dusoe}}, \bibinfo
		{author} {\bibfnamefont {K.~R.}\ \bibnamefont {Kittilstved}}, \bibinfo
		{author} {\bibfnamefont {D.}~\bibnamefont {Venkataraman}},\ and\ \bibinfo
		{author} {\bibfnamefont {Z.}~\bibnamefont {Ak{\v{s}}amija}},\ }\href@noop {}
	{\bibfield  {journal} {\bibinfo  {journal} {Adv. Sci.}\ }\textbf {\bibinfo
			{volume} {8}},\ \bibinfo {pages} {2101087} (\bibinfo {year}
		{2021})}\BibitemShut {NoStop}%
	\bibitem [{\citenamefont {Debye}\ and\ \citenamefont
		{Hückel}(1923)}]{Debye1923}%
	\BibitemOpen
	\bibfield  {author} {\bibinfo {author} {\bibfnamefont {P.}~\bibnamefont
			{Debye}}\ and\ \bibinfo {author} {\bibfnamefont {E.}~\bibnamefont
			{Hückel}},\ }\href@noop {} {\bibfield  {journal} {\bibinfo  {journal} {Phys.
				Z.}\ }\textbf {\bibinfo {volume} {24}} (\bibinfo {year} {1923})}\BibitemShut
	{NoStop}%
	\bibitem [{\citenamefont {Kjellander}(1995)}]{Kjellander1995}%
	\BibitemOpen
	\bibfield  {author} {\bibinfo {author} {\bibfnamefont {R.}~\bibnamefont
			{Kjellander}},\ }\href@noop {} {\bibfield  {journal} {\bibinfo  {journal} {J.
				Phys. Chem.}\ }\textbf {\bibinfo {volume} {99}},\ \bibinfo {pages} {10392}
		(\bibinfo {year} {1995})}\BibitemShut {NoStop}%
	\bibitem [{\citenamefont {Garavelli}\ and\ \citenamefont
		{Oliveira}(1991)}]{Garavelli1991}%
	\BibitemOpen
	\bibfield  {author} {\bibinfo {author} {\bibfnamefont {S.~L.}\ \bibnamefont
			{Garavelli}}\ and\ \bibinfo {author} {\bibfnamefont {F.~A.}\ \bibnamefont
			{Oliveira}},\ }\href@noop {} {\bibfield  {journal} {\bibinfo  {journal}
			{Phys. Rev. Lett.}\ }\textbf {\bibinfo {volume} {66}},\ \bibinfo {pages}
		{1310} (\bibinfo {year} {1991})}\BibitemShut {NoStop}%
	\bibitem [{\citenamefont {Napsuciale}\ and\ \citenamefont
		{Rodr{\'{\i}}guez}(2021)}]{Napsuciale2021}%
	\BibitemOpen
	\bibfield  {author} {\bibinfo {author} {\bibfnamefont {M.}~\bibnamefont
			{Napsuciale}}\ and\ \bibinfo {author} {\bibfnamefont {S.}~\bibnamefont
			{Rodr{\'{\i}}guez}},\ }\href@noop {} {\bibfield  {journal} {\bibinfo
			{journal} {Phys. Lett. B}\ }\textbf {\bibinfo {volume} {816}},\ \bibinfo
		{pages} {136218} (\bibinfo {year} {2021})}\BibitemShut {NoStop}%
	\bibitem [{\citenamefont {Yukawa}(1955)}]{Yukawa1955}%
	\BibitemOpen
	\bibfield  {author} {\bibinfo {author} {\bibfnamefont {H.}~\bibnamefont
			{Yukawa}},\ }\href@noop {} {\bibfield  {journal} {\bibinfo  {journal} {Prog.
				Theor. Phys. Supp.}\ }\textbf {\bibinfo {volume} {1}},\ \bibinfo {pages} {1}
		(\bibinfo {year} {1955})}\BibitemShut {NoStop}%
	\bibitem [{\citenamefont {Attard}(1993)}]{Attard1993}%
	\BibitemOpen
	\bibfield  {author} {\bibinfo {author} {\bibfnamefont {P.}~\bibnamefont
			{Attard}},\ }\href@noop {} {\bibfield  {journal} {\bibinfo  {journal} {Phys.
				Rev. E}\ }\textbf {\bibinfo {volume} {48}},\ \bibinfo {pages} {3604}
		(\bibinfo {year} {1993})}\BibitemShut {NoStop}%
	\bibitem [{\citenamefont {Rotenberg}\ \emph {et~al.}(2018)\citenamefont
		{Rotenberg}, \citenamefont {Bernard},\ and\ \citenamefont
		{Hansen}}]{Rotenberg2018}%
	\BibitemOpen
	\bibfield  {author} {\bibinfo {author} {\bibfnamefont {B.}~\bibnamefont
			{Rotenberg}}, \bibinfo {author} {\bibfnamefont {O.}~\bibnamefont {Bernard}},\
		and\ \bibinfo {author} {\bibfnamefont {J.-P.}\ \bibnamefont {Hansen}},\
	}\href@noop {} {\bibfield  {journal} {\bibinfo  {journal} {J. Phys.: Condens.
				Matter}\ }\textbf {\bibinfo {volume} {30}},\ \bibinfo {pages} {054005}
		(\bibinfo {year} {2018})}\BibitemShut {NoStop}%
	\bibitem [{\citenamefont {Krucker-Velasquez}\ and\ \citenamefont
		{Swan}(2021)}]{KruckerVelasquez2021}%
	\BibitemOpen
	\bibfield  {author} {\bibinfo {author} {\bibfnamefont {E.}~\bibnamefont
			{Krucker-Velasquez}}\ and\ \bibinfo {author} {\bibfnamefont {J.~W.}\
			\bibnamefont {Swan}},\ }\href@noop {} {\bibfield  {journal} {\bibinfo
			{journal} {J. Chem. Phys.}\ }\textbf {\bibinfo {volume} {155}},\ \bibinfo
		{pages} {134903} (\bibinfo {year} {2021})}\BibitemShut {NoStop}%
	\bibitem [{\citenamefont {Oelerich}\ \emph {et~al.}(2012)\citenamefont
		{Oelerich}, \citenamefont {Huemmer},\ and\ \citenamefont
		{Baranovskii}}]{Oelerich2012}%
	\BibitemOpen
	\bibfield  {author} {\bibinfo {author} {\bibfnamefont {J.~O.}\ \bibnamefont
			{Oelerich}}, \bibinfo {author} {\bibfnamefont {D.}~\bibnamefont {Huemmer}},\
		and\ \bibinfo {author} {\bibfnamefont {S.~D.}\ \bibnamefont {Baranovskii}},\
	}\href@noop {} {\bibfield  {journal} {\bibinfo  {journal} {Phys. Rev. Lett.}\
		}\textbf {\bibinfo {volume} {108}},\ \bibinfo {pages} {226403} (\bibinfo
		{year} {2012})}\BibitemShut {NoStop}%
	\bibitem [{\citenamefont {Debesay}\ \emph {et~al.}(2021)\citenamefont
		{Debesay}, \citenamefont {Sun},\ and\ \citenamefont
		{Bahoura}}]{DebesayAIMS21}%
	\BibitemOpen
	\bibfield  {author} {\bibinfo {author} {\bibfnamefont {T.}~\bibnamefont
			{Debesay}}, \bibinfo {author} {\bibfnamefont {S.-S.}\ \bibnamefont {Sun}},\
		and\ \bibinfo {author} {\bibfnamefont {M.}~\bibnamefont {Bahoura}},\
	}\href@noop {} {\bibfield  {journal} {\bibinfo  {journal} {AIMS Materials
				Science}\ }\textbf {\bibinfo {volume} {8}},\ \bibinfo {pages} {823} (\bibinfo
		{year} {2021})}\BibitemShut {NoStop}%
	\bibitem [{Iod()}]{Iodine}%
	\BibitemOpen
	\href@noop {} {\emph {\bibinfo {title} {See Supplemental Material [URL] for
				details on iodine IP and carried out measurement with procedure described in
				Ref. \cite{Ko2012}}}}\BibitemShut {NoStop}%
	\bibitem [{\citenamefont {Irwin}\ \emph {et~al.}(2008)\citenamefont {Irwin},
		\citenamefont {Buchholz}, \citenamefont {Hains}, \citenamefont {Chang},\ and\
		\citenamefont {Marks}}]{IrwinPNAS08}%
	\BibitemOpen
	\bibfield  {author} {\bibinfo {author} {\bibfnamefont {M.~D.}\ \bibnamefont
			{Irwin}}, \bibinfo {author} {\bibfnamefont {D.~B.}\ \bibnamefont {Buchholz}},
		\bibinfo {author} {\bibfnamefont {A.~W.}\ \bibnamefont {Hains}}, \bibinfo
		{author} {\bibfnamefont {R.~P.~H.}\ \bibnamefont {Chang}},\ and\ \bibinfo
		{author} {\bibfnamefont {T.~J.}\ \bibnamefont {Marks}},\ }\href@noop {}
	{\bibfield  {journal} {\bibinfo  {journal} {Proc. Natl. Acad. Sci.}\ }\textbf
		{\bibinfo {volume} {105}},\ \bibinfo {pages} {2783} (\bibinfo {year}
		{2008})}\BibitemShut {NoStop}%
	\bibitem [{\citenamefont {Fediai}\ \emph {et~al.}(2020)\citenamefont {Fediai},
		\citenamefont {Emering}, \citenamefont {Symalla},\ and\ \citenamefont
		{Wenzel}}]{FediaiPCCP20}%
	\BibitemOpen
	\bibfield  {author} {\bibinfo {author} {\bibfnamefont {A.}~\bibnamefont
			{Fediai}}, \bibinfo {author} {\bibfnamefont {A.}~\bibnamefont {Emering}},
		\bibinfo {author} {\bibfnamefont {F.}~\bibnamefont {Symalla}},\ and\ \bibinfo
		{author} {\bibfnamefont {W.}~\bibnamefont {Wenzel}},\ }\href@noop {}
	{\bibfield  {journal} {\bibinfo  {journal} {Phys. Chem. Chem. Phys.}\
		}\textbf {\bibinfo {volume} {22}},\ \bibinfo {pages} {10256} (\bibinfo {year}
		{2020})}\BibitemShut {NoStop}%
	\bibitem [{\citenamefont {Ukai}\ \emph {et~al.}(2005)\citenamefont {Ukai},
		\citenamefont {Ito}, \citenamefont {Marumoto},\ and\ \citenamefont {ichi
			Kuroda}}]{Ukai2005}%
	\BibitemOpen
	\bibfield  {author} {\bibinfo {author} {\bibfnamefont {S.}~\bibnamefont
			{Ukai}}, \bibinfo {author} {\bibfnamefont {H.}~\bibnamefont {Ito}}, \bibinfo
		{author} {\bibfnamefont {K.}~\bibnamefont {Marumoto}},\ and\ \bibinfo
		{author} {\bibfnamefont {S.}~\bibnamefont {ichi Kuroda}},\ }\href@noop {}
	{\bibfield  {journal} {\bibinfo  {journal} {J. Phys. Soc. Japan}\ }\textbf
		{\bibinfo {volume} {74}},\ \bibinfo {pages} {3314} (\bibinfo {year}
		{2005})}\BibitemShut {NoStop}%
	\bibitem [{\citenamefont {Jacobs}\ \emph {et~al.}(2022)\citenamefont {Jacobs},
		\citenamefont {D'Avino}, \citenamefont {Lemaur}, \citenamefont {Lin},
		\citenamefont {Huang}, \citenamefont {Chen}, \citenamefont {Harrelson},
		\citenamefont {Wood}, \citenamefont {Spalek}, \citenamefont {Mustafa},
		\citenamefont {O'Keefe}, \citenamefont {Ren}, \citenamefont {Simatos},
		\citenamefont {Tjhe}, \citenamefont {Statz}, \citenamefont {Strzalka},
		\citenamefont {Lee}, \citenamefont {McCulloch}, \citenamefont {Fratini},
		\citenamefont {Beljonne},\ and\ \citenamefont {Sirringhaus}}]{Jacobs2022}%
	\BibitemOpen
	\bibfield  {author} {\bibinfo {author} {\bibfnamefont {I.~E.}\ \bibnamefont
			{Jacobs}}, \bibinfo {author} {\bibfnamefont {G.}~\bibnamefont {D'Avino}},
		\bibinfo {author} {\bibfnamefont {V.}~\bibnamefont {Lemaur}}, \bibinfo
		{author} {\bibfnamefont {Y.}~\bibnamefont {Lin}}, \bibinfo {author}
		{\bibfnamefont {Y.}~\bibnamefont {Huang}}, \bibinfo {author} {\bibfnamefont
			{C.}~\bibnamefont {Chen}}, \bibinfo {author} {\bibfnamefont {T.~F.}\
			\bibnamefont {Harrelson}}, \bibinfo {author} {\bibfnamefont {W.}~\bibnamefont
			{Wood}}, \bibinfo {author} {\bibfnamefont {L.~J.}\ \bibnamefont {Spalek}},
		\bibinfo {author} {\bibfnamefont {T.}~\bibnamefont {Mustafa}}, \bibinfo
		{author} {\bibfnamefont {C.~A.}\ \bibnamefont {O'Keefe}}, \bibinfo {author}
		{\bibfnamefont {X.}~\bibnamefont {Ren}}, \bibinfo {author} {\bibfnamefont
			{D.}~\bibnamefont {Simatos}}, \bibinfo {author} {\bibfnamefont
			{D.}~\bibnamefont {Tjhe}}, \bibinfo {author} {\bibfnamefont {M.}~\bibnamefont
			{Statz}}, \bibinfo {author} {\bibfnamefont {J.~W.}\ \bibnamefont {Strzalka}},
		\bibinfo {author} {\bibfnamefont {J.-K.}\ \bibnamefont {Lee}}, \bibinfo
		{author} {\bibfnamefont {I.}~\bibnamefont {McCulloch}}, \bibinfo {author}
		{\bibfnamefont {S.}~\bibnamefont {Fratini}}, \bibinfo {author} {\bibfnamefont
			{D.}~\bibnamefont {Beljonne}},\ and\ \bibinfo {author} {\bibfnamefont
			{H.}~\bibnamefont {Sirringhaus}},\ }\href@noop {} {\bibfield  {journal}
		{\bibinfo  {journal} {J. Am. Chem. Soc.}\ }\textbf {\bibinfo {volume}
			{144}},\ \bibinfo {pages} {3005} (\bibinfo {year} {2022})}\BibitemShut
	{NoStop}%
	\bibitem [{\citenamefont {Suh}\ \emph {et~al.}(2019)\citenamefont {Suh},
		\citenamefont {Jeong}, \citenamefont {Oh}, \citenamefont {Lee}, \citenamefont
		{Jung}, \citenamefont {Kang},\ and\ \citenamefont {Jang}}]{Suh2019}%
	\BibitemOpen
	\bibfield  {author} {\bibinfo {author} {\bibfnamefont {E.~H.}\ \bibnamefont
			{Suh}}, \bibinfo {author} {\bibfnamefont {Y.~J.}\ \bibnamefont {Jeong}},
		\bibinfo {author} {\bibfnamefont {J.~G.}\ \bibnamefont {Oh}}, \bibinfo
		{author} {\bibfnamefont {K.}~\bibnamefont {Lee}}, \bibinfo {author}
		{\bibfnamefont {J.}~\bibnamefont {Jung}}, \bibinfo {author} {\bibfnamefont
			{Y.~S.}\ \bibnamefont {Kang}},\ and\ \bibinfo {author} {\bibfnamefont
			{J.}~\bibnamefont {Jang}},\ }\href@noop {} {\bibfield  {journal} {\bibinfo
			{journal} {Nano Energy}\ }\textbf {\bibinfo {volume} {58}},\ \bibinfo {pages}
		{585} (\bibinfo {year} {2019})}\BibitemShut {NoStop}%
	\bibitem [{\citenamefont {Hynynen}\ \emph {et~al.}(2017)\citenamefont
		{Hynynen}, \citenamefont {Kiefer}, \citenamefont {Yu}, \citenamefont {Kroon},
		\citenamefont {Munir}, \citenamefont {Amassian}, \citenamefont {Kemerink},\
		and\ \citenamefont {Müller}}]{Hynynen2017}%
	\BibitemOpen
	\bibfield  {author} {\bibinfo {author} {\bibfnamefont {J.}~\bibnamefont
			{Hynynen}}, \bibinfo {author} {\bibfnamefont {D.}~\bibnamefont {Kiefer}},
		\bibinfo {author} {\bibfnamefont {L.}~\bibnamefont {Yu}}, \bibinfo {author}
		{\bibfnamefont {R.}~\bibnamefont {Kroon}}, \bibinfo {author} {\bibfnamefont
			{R.}~\bibnamefont {Munir}}, \bibinfo {author} {\bibfnamefont
			{A.}~\bibnamefont {Amassian}}, \bibinfo {author} {\bibfnamefont
			{M.}~\bibnamefont {Kemerink}},\ and\ \bibinfo {author} {\bibfnamefont
			{C.}~\bibnamefont {Müller}},\ }\href@noop {} {\bibfield  {journal} {\bibinfo
			{journal} {Macromolecules}\ }\textbf {\bibinfo {volume} {50}},\ \bibinfo
		{pages} {8140} (\bibinfo {year} {2017})}\BibitemShut {NoStop}%
	\bibitem [{\citenamefont {Hamidi-Sakr}\ \emph {et~al.}(2017)\citenamefont
		{Hamidi-Sakr}, \citenamefont {Biniek}, \citenamefont {Bantignies},
		\citenamefont {Maurin}, \citenamefont {Herrmann}, \citenamefont {Leclerc},
		\citenamefont {L{\'{e}}v{\^{e}}que}, \citenamefont {Vijayakumar},
		\citenamefont {Zimmermann},\ and\ \citenamefont
		{Brinkmann}}]{HamidiSakr2017}%
	\BibitemOpen
	\bibfield  {author} {\bibinfo {author} {\bibfnamefont {A.}~\bibnamefont
			{Hamidi-Sakr}}, \bibinfo {author} {\bibfnamefont {L.}~\bibnamefont {Biniek}},
		\bibinfo {author} {\bibfnamefont {J.-L.}\ \bibnamefont {Bantignies}},
		\bibinfo {author} {\bibfnamefont {D.}~\bibnamefont {Maurin}}, \bibinfo
		{author} {\bibfnamefont {L.}~\bibnamefont {Herrmann}}, \bibinfo {author}
		{\bibfnamefont {N.}~\bibnamefont {Leclerc}}, \bibinfo {author} {\bibfnamefont
			{P.}~\bibnamefont {L{\'{e}}v{\^{e}}que}}, \bibinfo {author} {\bibfnamefont
			{V.}~\bibnamefont {Vijayakumar}}, \bibinfo {author} {\bibfnamefont
			{N.}~\bibnamefont {Zimmermann}},\ and\ \bibinfo {author} {\bibfnamefont
			{M.}~\bibnamefont {Brinkmann}},\ }\href@noop {} {\bibfield  {journal}
		{\bibinfo  {journal} {Adv. Funct. Mater.}\ }\textbf {\bibinfo {volume}
			{27}},\ \bibinfo {pages} {1700173} (\bibinfo {year} {2017})}\BibitemShut
	{NoStop}%
	\bibitem [{\citenamefont {Vijayakumar}\ \emph {et~al.}(2019)\citenamefont
		{Vijayakumar}, \citenamefont {Zaborova}, \citenamefont {Biniek},
		\citenamefont {Zeng}, \citenamefont {Herrmann}, \citenamefont {Carvalho},
		\citenamefont {Boyron}, \citenamefont {Leclerc},\ and\ \citenamefont
		{Brinkmann}}]{Vijayakumar2019}%
	\BibitemOpen
	\bibfield  {author} {\bibinfo {author} {\bibfnamefont {V.}~\bibnamefont
			{Vijayakumar}}, \bibinfo {author} {\bibfnamefont {E.}~\bibnamefont
			{Zaborova}}, \bibinfo {author} {\bibfnamefont {L.}~\bibnamefont {Biniek}},
		\bibinfo {author} {\bibfnamefont {H.}~\bibnamefont {Zeng}}, \bibinfo {author}
		{\bibfnamefont {L.}~\bibnamefont {Herrmann}}, \bibinfo {author}
		{\bibfnamefont {A.}~\bibnamefont {Carvalho}}, \bibinfo {author}
		{\bibfnamefont {O.}~\bibnamefont {Boyron}}, \bibinfo {author} {\bibfnamefont
			{N.}~\bibnamefont {Leclerc}},\ and\ \bibinfo {author} {\bibfnamefont
			{M.}~\bibnamefont {Brinkmann}},\ }\href@noop {} {\bibfield  {journal}
		{\bibinfo  {journal} {ACS Appl. Mater. Interfaces}\ }\textbf {\bibinfo
			{volume} {11}},\ \bibinfo {pages} {4942} (\bibinfo {year}
		{2019})}\BibitemShut {NoStop}%
	\bibitem [{\citenamefont {Fediai}\ \emph {et~al.}(2019)\citenamefont {Fediai},
		\citenamefont {Symalla}, \citenamefont {Friederich},\ and\ \citenamefont
		{Wenzel}}]{Fediai2019}%
	\BibitemOpen
	\bibfield  {author} {\bibinfo {author} {\bibfnamefont {A.}~\bibnamefont
			{Fediai}}, \bibinfo {author} {\bibfnamefont {F.}~\bibnamefont {Symalla}},
		\bibinfo {author} {\bibfnamefont {P.}~\bibnamefont {Friederich}},\ and\
		\bibinfo {author} {\bibfnamefont {W.}~\bibnamefont {Wenzel}},\ }\href@noop {}
	{\bibfield  {journal} {\bibinfo  {journal} {Nature Commun.}\ }\textbf
		{\bibinfo {volume} {10}} (\bibinfo {year} {2019})}\BibitemShut {NoStop}%
	\bibitem [{\citenamefont {Warren}\ \emph {et~al.}(2023)\citenamefont {Warren},
		\citenamefont {Blom},\ and\ \citenamefont {Koch}}]{WarrenAPL23}%
	\BibitemOpen
	\bibfield  {author} {\bibinfo {author} {\bibfnamefont {R.}~\bibnamefont
			{Warren}}, \bibinfo {author} {\bibfnamefont {P.~W.~M.}\ \bibnamefont
			{Blom}},\ and\ \bibinfo {author} {\bibfnamefont {N.}~\bibnamefont {Koch}},\
	}\href@noop {} {\bibfield  {journal} {\bibinfo  {journal} {Appl. Phys.
				Lett.}\ }\textbf {\bibinfo {volume} {122}},\ \bibinfo {pages} {152108}
		(\bibinfo {year} {2023})}\BibitemShut {NoStop}%
	\bibitem [{\citenamefont {Kramer}\ \emph {et~al.}(2022)\citenamefont {Kramer},
		\citenamefont {Kaiser}, \citenamefont {Zhang}, \citenamefont {Murthy},
		\citenamefont {Gagliardi}, \citenamefont {Hsu},\ and\ \citenamefont
		{Vandenberghe}}]{KramerNanoEner22}%
	\BibitemOpen
	\bibfield  {author} {\bibinfo {author} {\bibfnamefont {A.}~\bibnamefont
			{Kramer}}, \bibinfo {author} {\bibfnamefont {W.}~\bibnamefont {Kaiser}},
		\bibinfo {author} {\bibfnamefont {B.}~\bibnamefont {Zhang}}, \bibinfo
		{author} {\bibfnamefont {L.~N.}\ \bibnamefont {Murthy}}, \bibinfo {author}
		{\bibfnamefont {A.}~\bibnamefont {Gagliardi}}, \bibinfo {author}
		{\bibfnamefont {J.~W.}\ \bibnamefont {Hsu}},\ and\ \bibinfo {author}
		{\bibfnamefont {W.~G.}\ \bibnamefont {Vandenberghe}},\ }\href@noop {}
	{\bibfield  {journal} {\bibinfo  {journal} {Nano Energy}\ }\textbf {\bibinfo
			{volume} {103}},\ \bibinfo {pages} {107793} (\bibinfo {year}
		{2022})}\BibitemShut {NoStop}%
	\bibitem [{\citenamefont {Zhao}\ \emph {et~al.}(2021)\citenamefont {Zhao},
		\citenamefont {Tang}, \citenamefont {Seah}, \citenamefont {Koh},
		\citenamefont {Chua}, \citenamefont {Png},\ and\ \citenamefont
		{Ho}}]{Zhao2021}%
	\BibitemOpen
	\bibfield  {author} {\bibinfo {author} {\bibfnamefont {C.}~\bibnamefont
			{Zhao}}, \bibinfo {author} {\bibfnamefont {C.~G.}\ \bibnamefont {Tang}},
		\bibinfo {author} {\bibfnamefont {Z.-L.}\ \bibnamefont {Seah}}, \bibinfo
		{author} {\bibfnamefont {Q.-M.}\ \bibnamefont {Koh}}, \bibinfo {author}
		{\bibfnamefont {L.-L.}\ \bibnamefont {Chua}}, \bibinfo {author}
		{\bibfnamefont {R.-Q.}\ \bibnamefont {Png}},\ and\ \bibinfo {author}
		{\bibfnamefont {P.~K.~H.}\ \bibnamefont {Ho}},\ }\href@noop {} {\bibfield
		{journal} {\bibinfo  {journal} {Nat. Commun.}\ }\textbf {\bibinfo {volume}
			{12}} (\bibinfo {year} {2021})}\BibitemShut {NoStop}%
	\bibitem [{\citenamefont {Zuo}\ \emph {et~al.}(2016)\citenamefont {Zuo},
		\citenamefont {Abdalla},\ and\ \citenamefont {Kemerink}}]{Zuo2016}%
	\BibitemOpen
	\bibfield  {author} {\bibinfo {author} {\bibfnamefont {G.}~\bibnamefont
			{Zuo}}, \bibinfo {author} {\bibfnamefont {H.}~\bibnamefont {Abdalla}},\ and\
		\bibinfo {author} {\bibfnamefont {M.}~\bibnamefont {Kemerink}},\ }\href
	{https://doi.org/10.1103/physrevb.93.235203} {\bibfield  {journal} {\bibinfo
			{journal} {Physical Review B}\ }\textbf {\bibinfo {volume} {93}},\ \bibinfo
		{pages} {235203} (\bibinfo {year} {2016})}\BibitemShut {NoStop}%
	\bibitem [{\citenamefont {Zuo}\ \emph {et~al.}(2018)\citenamefont {Zuo},
		\citenamefont {Abdalla},\ and\ \citenamefont {Kemerink}}]{Zuo2018}%
	\BibitemOpen
	\bibfield  {author} {\bibinfo {author} {\bibfnamefont {G.}~\bibnamefont
			{Zuo}}, \bibinfo {author} {\bibfnamefont {H.}~\bibnamefont {Abdalla}},\ and\
		\bibinfo {author} {\bibfnamefont {M.}~\bibnamefont {Kemerink}},\ }\href@noop
	{} {\bibfield  {journal} {\bibinfo  {journal} {Phys. Rev. B}\ }\textbf
		{\bibinfo {volume} {97}},\ \bibinfo {pages} {079902} (\bibinfo {year}
		{2018})}\BibitemShut {NoStop}%
	\bibitem [{\citenamefont {Ko}\ \emph {et~al.}(2012)\citenamefont {Ko},
		\citenamefont {Hoke}, \citenamefont {Pandey}, \citenamefont {Hong},
		\citenamefont {Mondal}, \citenamefont {Risko}, \citenamefont {Yi},
		\citenamefont {Noriega}, \citenamefont {McGehee}, \citenamefont
		{Br{\'{e}}das}, \citenamefont {Salleo},\ and\ \citenamefont {Bao}}]{Ko2012}%
	\BibitemOpen
	\bibfield  {author} {\bibinfo {author} {\bibfnamefont {S.}~\bibnamefont
			{Ko}}, \bibinfo {author} {\bibfnamefont {E.~T.}\ \bibnamefont {Hoke}},
		\bibinfo {author} {\bibfnamefont {L.}~\bibnamefont {Pandey}}, \bibinfo
		{author} {\bibfnamefont {S.}~\bibnamefont {Hong}}, \bibinfo {author}
		{\bibfnamefont {R.}~\bibnamefont {Mondal}}, \bibinfo {author} {\bibfnamefont
			{C.}~\bibnamefont {Risko}}, \bibinfo {author} {\bibfnamefont
			{Y.}~\bibnamefont {Yi}}, \bibinfo {author} {\bibfnamefont {R.}~\bibnamefont
			{Noriega}}, \bibinfo {author} {\bibfnamefont {M.~D.}\ \bibnamefont
			{McGehee}}, \bibinfo {author} {\bibfnamefont {J.-L.}\ \bibnamefont
			{Br{\'{e}}das}}, \bibinfo {author} {\bibfnamefont {A.}~\bibnamefont
			{Salleo}},\ and\ \bibinfo {author} {\bibfnamefont {Z.}~\bibnamefont {Bao}},\
	}\href {https://doi.org/10.1021/ja210954r} {\bibfield  {journal} {\bibinfo
			{journal} {Journal of the American Chemical Society}\ }\textbf {\bibinfo
			{volume} {134}},\ \bibinfo {pages} {5222} (\bibinfo {year}
		{2012})}\BibitemShut {NoStop}%
\end{thebibliography}
%

\end{document}